# Remote Sub-Wavelength Focusing of Ultrasonically Activated Lorentz Current

Angad S. Rekhi[a)] and Amin Arbabian
*Department of Electrical Engineering, Stanford University, Stanford, California 94305, USA*

We propose the use of a combination of ultrasonic and magnetic fields in conductive media for the creation of RF electrical current via the Lorentz force, in order to achieve current generation with extreme sub-wavelength resolution at large depth. We demonstrate the modeling, generation, and measurement of Lorentz current in conductive solution, and show that this current can be localized at a distance of 13 *cm* from the ultrasonic source to a region about three orders of magnitude smaller than the corresponding wavelength of electromagnetic waves at the same operation frequency. Our results exhibit greater depth, tighter localization, and closer agreement with prediction than previous work on the measurement of Lorentz current in a solution of homogeneous conductivity. The proposed method of RF current excitation overcomes the trade-off between focusing and propagation that is fundamental in the use of RF electromagnetic excitation alone, and has the potential to improve localization and depth of operation for RF current-based biomedical applications.

The problem of focusing waves at large depth is important in a variety of research areas, such as microscopy, seismology, and tomography. In particular, the ability to focus RF electromagnetic waves at depth in the body can increase the utility of a number of biomedical applications, including microwave imaging,[1] electrically controlled drug release,[2] electrical impedance tomography,[3,4] and non-invasive neural stimulation,[5,6] with the lattermost application increasing in importance as electroceuticals come to the fore.[7–9]

There is, however, a fundamental trade-off between propagation and focusing of RF electromagnetic waves in tissue: as the frequency of excitation is increased in order to improve spatial resolution, the tissue absorption increases as well.[10] This trade-off makes it difficult to achieve focused operation at large depth. If the emitted intensity is increased to compensate for the reduction in amplitude due to greater absorption, then the wave loses more energy to absorption during propagation, which can limit the efficiency and efficacy of the entire system.

One way to overcome the trade-off between propagation and focusing is to implement *transduction* at the region of interest. For instance, the release of *nm*-sized contrast agents enables sub-*mm* resolution for photoacoustic imaging,[11] while optogenetic methods can be used to selectively excite or inhibit neurons with single-cell resolution.[12] However, these approaches achieve high degrees of localization at the expense of invasiveness.

We propose the use of a combination of ultrasonic and magnetic fields for the generation of electrical current at large tissue depth. Figure 1 shows a conceptual overview of our method of transduction. In the presence of an incident ultrasonic wave, initially stationary charged particles oscillate along the direction of wave propagation. With the addition of a static magnetic field, the ultrasonically activated movement of the charged particles gives rise to motion in a direction normal to both the magnetic field and the direction of wave propagation, due to the Lorentz force. Since the direction of the Lorentz force is dependent on the sign of the charge, this method of excitation creates charge separation, which is measurable as electrical current. In a medium with bulk conductivity $\sigma$, this method of transduction can be succinctly captured by the following equation:

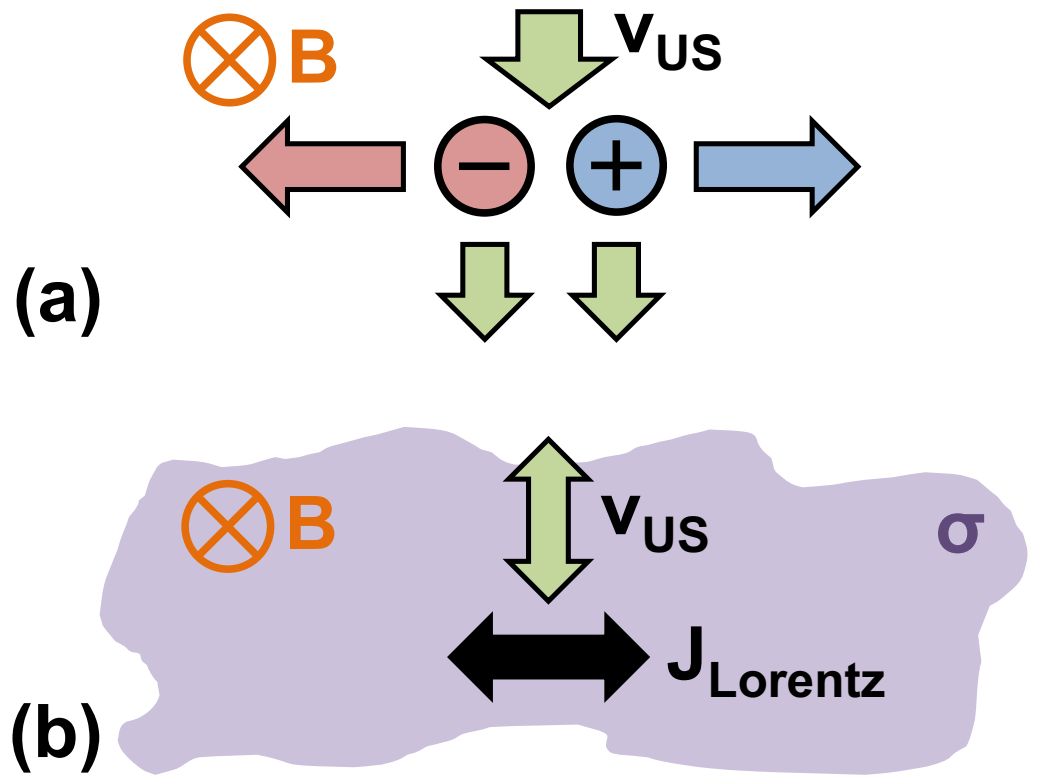

Figure 1: (a) Two-particle model demonstrating the Lorentz force-induced movement of ultrasonically excited charges in the presence of a magnetic field. In this diagram, the particles are assumed to not interact with each other, for simplicity. (b) Extension to a medium of bulk conductivity $\sigma$.

$$\mathbf{J_{Lorentz}} = \sigma\, \mathbf{v_{US}} \times \mathbf{B}, \tag{1}$$

where $\mathbf{v_{US}}$ is the ultrasonic velocity, $\mathbf{B}$ is the magnetic field, and $\mathbf{J_{Lorentz}}$ is the Lorentz current density.

With this method of transduction, the frequency of ultrasonic excitation can be chosen to minimize propagation loss, while the spatial parameters of the ultrasonic and magnetic fields can be chosen to shape the profile of the excited Lorentz current. Because the speed of sound in tissue is so much smaller than the speed of light, the focal size achievable using this method is orders of magnitude smaller than the wavelength of an electromagnetic

[a)]arekhi@ieee.org.

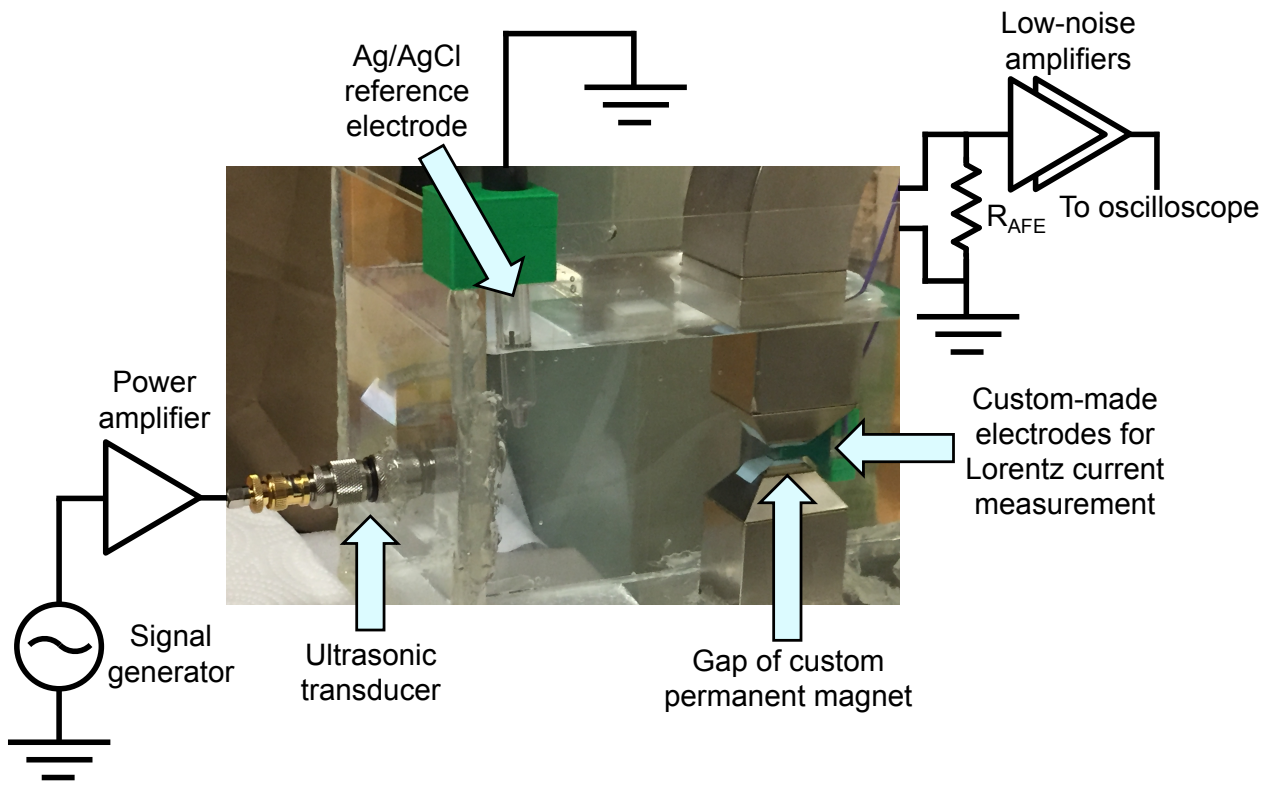

Figure 2: Experimental setup with system components labeled.

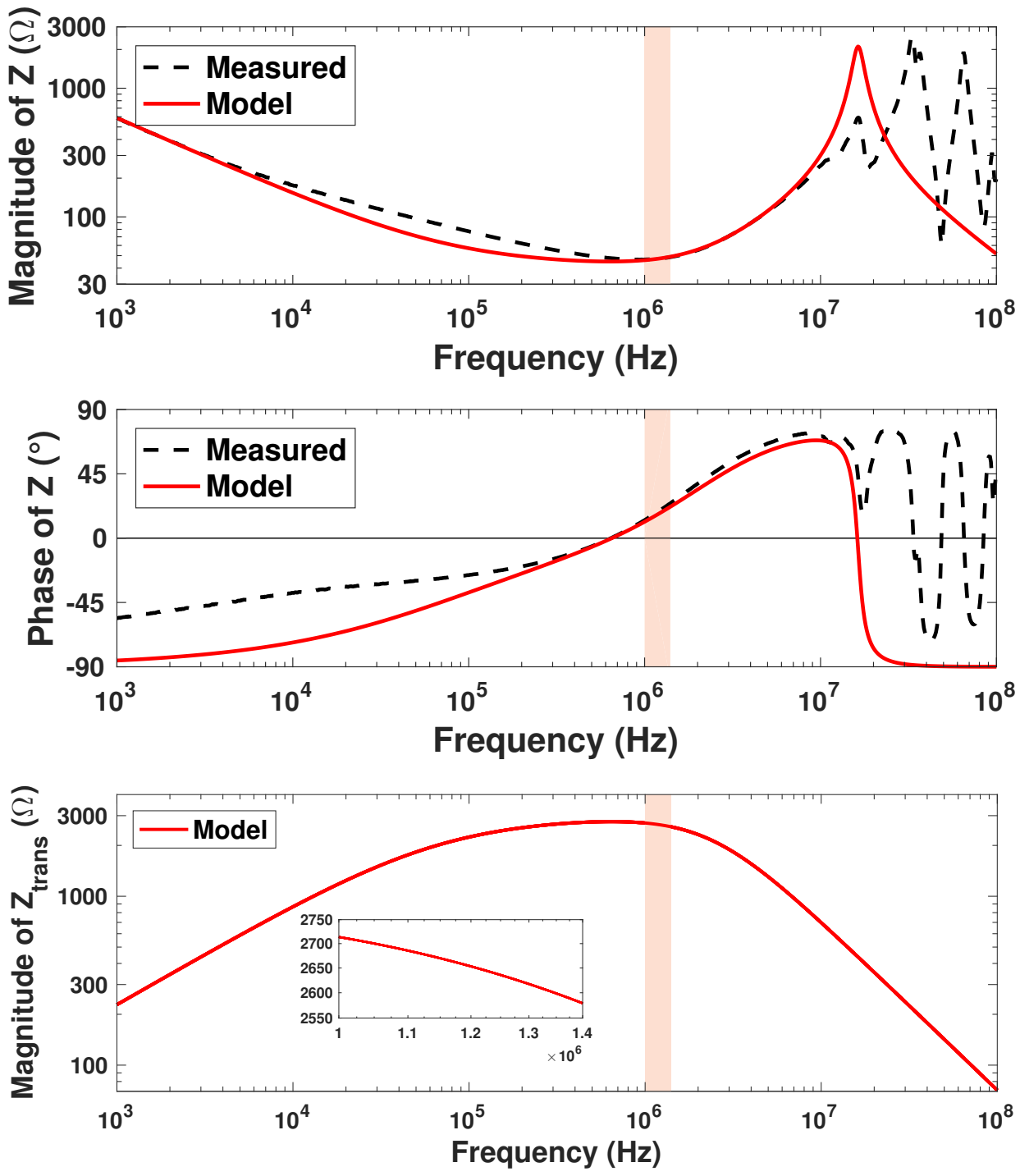

Figure 4: Top, center: Measured and modeled impedance of the electrodes in 1 M KCl solution. Maximum error between model and measurement in the bandwidth of interest is 1.2% (magnitude) and 2.5° (phase). Bottom: transimpedance $Z_{trans}$ of the signal chain, defined as the gain from the Lorentz current source to the voltage displayed on the oscilloscope.

wave at the same frequency, leading to sub-wavelength RF current excitation. The proposed method of Lorentz current generation therefore overcomes the trade-off between propagation and focusing that is inherent in the use of RF electromagnetic excitation alone, and does so without resorting to invasive techniques, as both the ultrasonic and magnetic fields can be applied from outside the body.

In this work, we demonstrate our proposed concept by modeling, generating, and measuring Lorentz current in ionic solution at a distance of 13 $cm$ from an unfocused ultrasonic transducer. The frequency of the generated current is near 1 $MHz$ while the majority of the current is spatially confined by the magnetic field profile, magnet gap, and measurement electrodes to a region about 1 $cm$ on a side, which is approximately $3 \times 10^3$ times smaller than the electromagnetic wavelength in water.

Using a focused ultrasonic transducer or transducer array would allow confinement of the generated Lorentz current to within an acoustic wavelength, which is about $2 \times 10^4$ times smaller than the corresponding electromagnetic wavelength. Using the spatial profile of the magnetic field to further confine the generated current (by, for example, offsetting the spatial peak of the magnetic field from the acoustic focus) would allow even tighter localization of the current, beyond that afforded by the ratio of the speeds of light and sound in tissue.

Compared to related prior work, we generate current with greater localization and at a larger distance from the ultrasonic source, which is promising for extending the depth and resolution of biomedical applications involving electrical current. Additionally, by precisely designing and modeling the excitation and measurement apparatus, we arrive at closer quantitative agreement between prediction and measurement.[13–15]

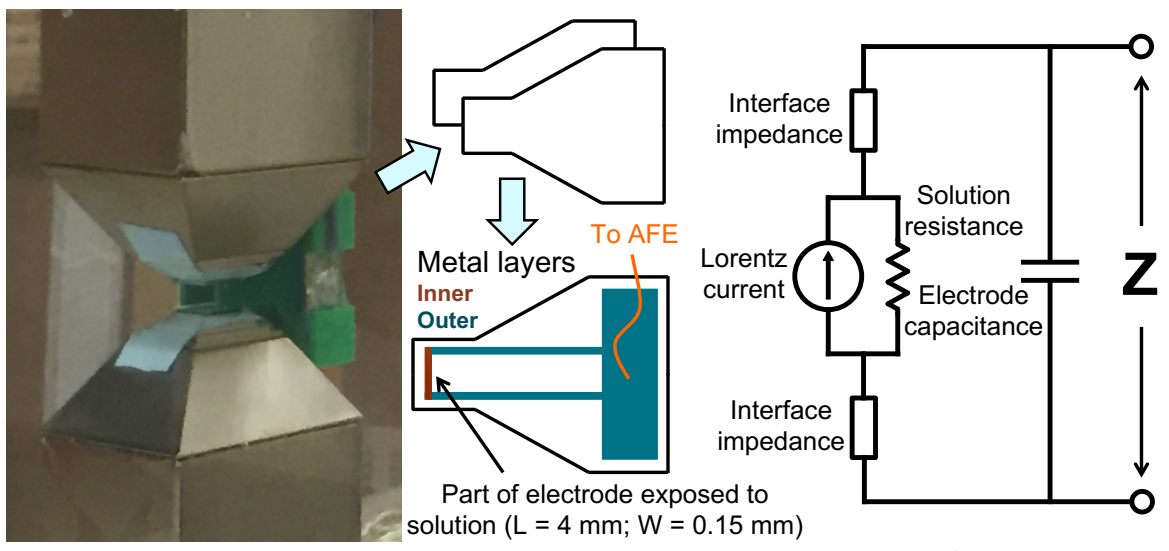

Figure 3: Layout and equivalent circuit model of electrodes.

Our experimental procedure is as follows: we send a pulsed ultrasonic wave into a region within an ionic solution with a strong magnetic field, monitor the current generated in that region using electrodes whose design (discussed below) is tailored to the task of current measurement in the chosen bandwidth, and use time-of-flight (TOF) to identify the expected Lorentz current.

We now describe our setup, which is shown in Figure 2, in more detail. We use a signal generator (*Keysight 33511B*) and broadband RF power amplifier (*E&I 411LA*) to excite an unfocused ultrasonic transducer (*Olympus A303S*) with a 30-cycle pulse of varying amplitude (20–90 $V_{pp}$; corresponding to a pressure range of 90–420 $kPa_{pp}$ at 13 $cm$ from the transducer, at 1.1 MHz) and frequency (1–1.4 $MHz$). The transducer is mounted on the side wall of a tank containing KCl solution, whose concentration is varied over several experiments in order to vary its conductivity in the range of 0.03–12 $S/m$ (*Malvern ZetaSizer Nano ZS90*).

The transducer is positioned so that the pulsed ultrasonic signal propagates a distance of 13 $cm$ into the gap of a custom-ordered permanent magnet (*SuperMagnetMan*); in the gap, the magnetic field strength is 2 $T$, and the field lines span the gap, running from one face of the magnet to the other. According to the Lorentz current hypothesis, the propagation of the ultrasonic wave through the gap of the magnet creates a current density $\boldsymbol{J}$ whose characteristics in time (frequency, pulse width) match those of the ultrasonic wave and whose direction is normal to the direction of propagation of the ultrasonic wave and to the direction of the magnetic field.

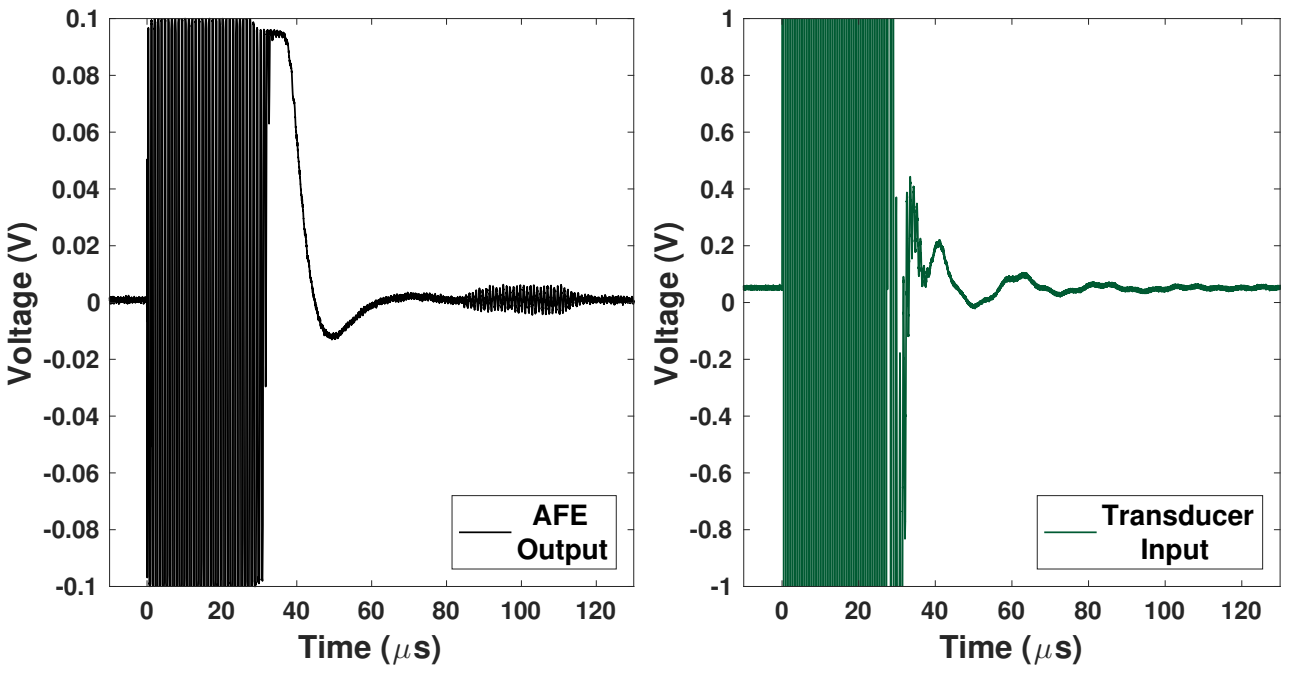

Figure 5: Measurements for a 30-cycle 1.1 $MHz$ pulse of amplitude 90.5 $V_{pp}$ (across ultrasonic transducer) emitted into 1 M KCl solution. Left: output of AFE. Right: input to transducer.

This current density $\boldsymbol{J}$ is converted into a current via measurement with two mirrored (but otherwise identical) custom-made electrodes that comprise printed metal traces on a circuit board; a close-up is shown in Figure 3. To maximize the amplitude of the measured current, the electrodes are made as long as possible (4 $mm$) while still fitting in the magnet gap; to minimize the averaging of current over an ultrasonic cycle, the width of the electrodes is 0.15 $mm$, which is about seven times smaller than the smallest wavelength of interest.

In order to guide the design of the analog front-end (AFE) connected to the electrodes, we measured the impedance of the electrodes in 1 M KCl solution ($\sigma = 11.7$ $S/m$); the circuit model of the electrodes is shown in Figure 3. The model consists of a solution resistance, an impedance that models the double-layer capacitance and other dynamics at each solution-metal interface, and a shunt capacitance that sits in parallel with the entire series network. This model captures the essential features of the impedance of the electrodes over the bandwidth of interest, as shown in Figure 4. Because the Lorentz current is generated in solution, it is modeled using a current source in parallel with the solution resistance.

Using the measured impedance of the electrodes, we designed the AFE for a transimpedance ($Z_{trans}$) of 2.6–2.7 $k\Omega$ over the bandwidth of interest, as shown in Figure 4, using $R_{AFE} = 3.8$ $\Omega$ and two low-noise amplifiers (*MiniCircuits ZFL-1000LN+*). The noise was calculated to be dominated by the solution resistance. Future work includes exploring the trade-off between filtering for noise reduction and Lorentz current pulse shape fidelity.

We used a Ag/AgCl reference electrode (*Aldrich Z113107*) to set the potential of the solution to the system ground. Here, since the inputs to the low-noise amplifiers in the AFE are AC-coupled, this was not strictly necessary; if a DC-coupled AFE were used, a reference electrode would be required, and its placement relative to the custom-made electrodes measuring the Lorentz currents would be an important design parameter.

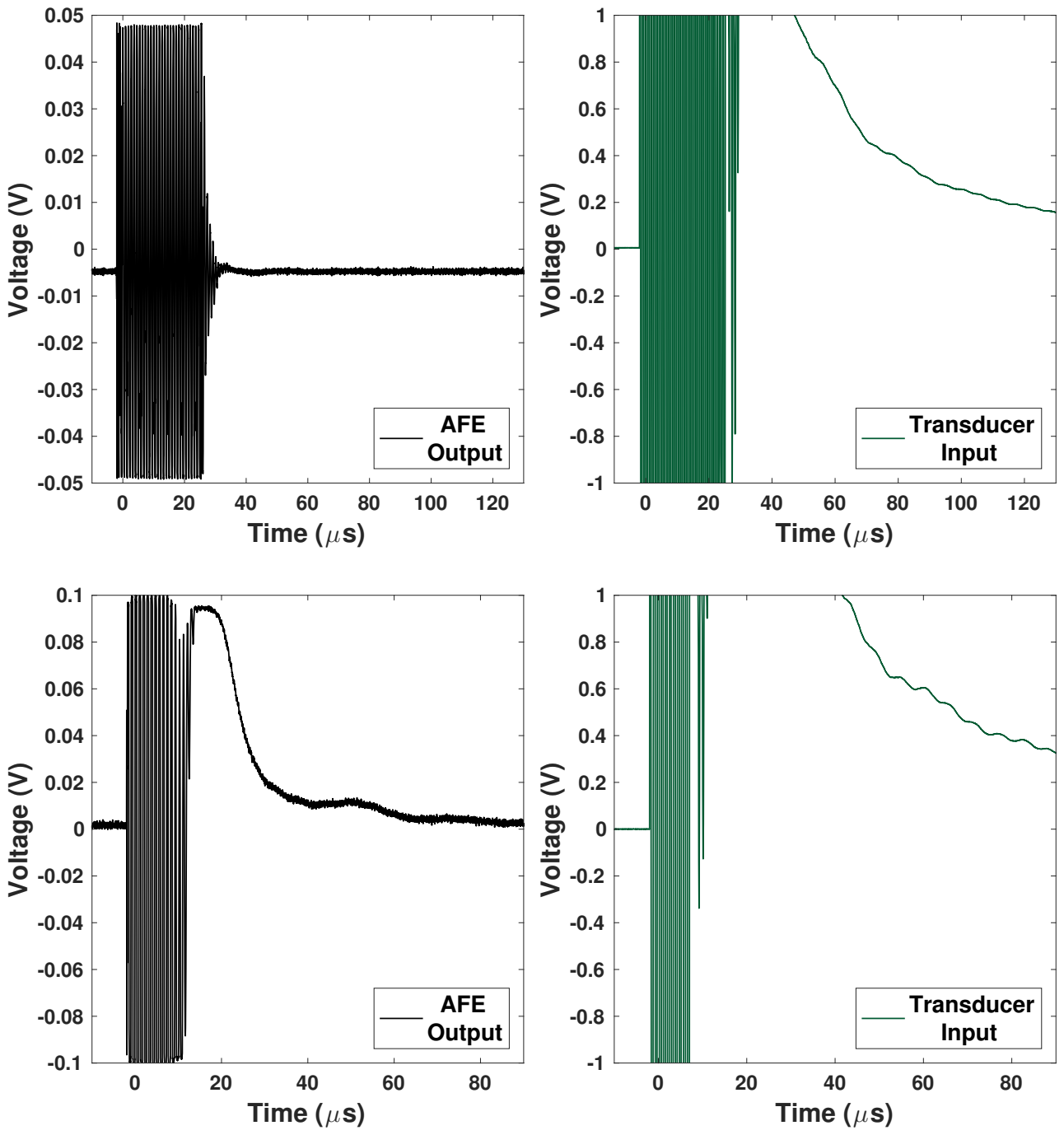

Figure 6: Control measurements. Top: KCl solution replaced by water ($\sigma = 0.03$ $S/m$); all other parameters identical to experiment whose results are shown in Figure 5. Bottom: electrodes moved to a location with $|\mathbf{B}| < 0.04$ $T$; pulse width shortened to ten cycles, but all other parameters identical to experiment whose results are shown in Figure 5.

We recorded both the output of the AFE and the input to the transducer for various combinations of ultrasonic frequency and amplitude and solution conductivity. To acquaint the reader with the nature of the results obtained from our experimental setup, measurements for a nominal case of these parameters are shown in Figure 5.

The large pulse measured at the AFE output for the first 30 $\mu s$ after transducer excitation is due to electromagnetic coupling from the cable at the output of the power amplifier to the wires running from the electrodes to the AFE. Subsequent ringing (30–70 $\mu s$) is due to interaction between the pulse and the electrode impedance.

The pulse of smaller amplitude starting at 85 $\mu s$ is a direct measurement of the hypothesized Lorentz current: its arrival time is consistent with prediction for a distance of 13 $cm$ at a speed of 1484 $m/s$, and its amplitude is close to the predicted value, as discussed below. Not shown in these plots are the pulses that reflect off the faces of the magnet and tank, excite the transducer and are converted into electrical signals, and are then coupled to the AFE output by the same mechanism as for the initial transducer excitation. Because these echo signals originate from waves that must travel at least twice as far as the wave that excites the Lorentz current, they are easily discriminated based on round-trip TOF.

To verify the physical origin of the observed signal at 85 $\mu s$, we conducted two control experiments. First, we replaced the KCl solution with water ($\sigma = 0.03$ $S/m$); the results are shown at the top of Figure 6. We observed no signal at 85 $\mu s$, the time at which a signal would be expected based on TOF considerations, indicating a successful negative control. For this control, the excitation signal that is electromagnetically coupled into the AFE has a different magnitude and dynamic response than for the nominal case shown in Figure 5, likely because the electrode-solution interface, which shapes this parasitic signal, now has a different impedance.

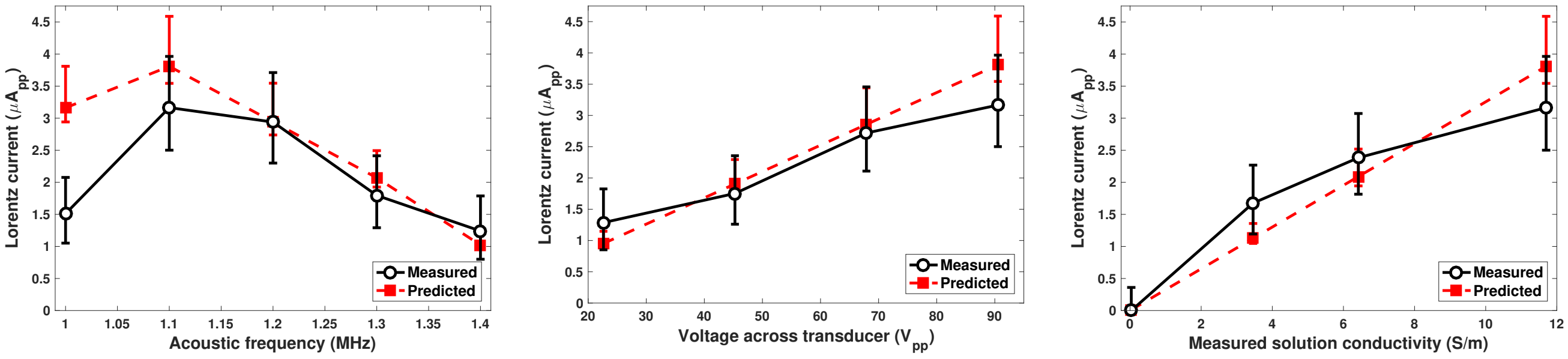

Figure 7: Comparison of measured Lorentz current amplitude to prediction across ultrasonic frequency (left) and amplitude (center) and solution conductivity (right). Sources of error for predicted values include uncertainty in magnetic field strength and in solution conductivity. Sources of error for measured values include voltage noise and uncertainty in transimpedance gain. Horizontal error bars on the rightmost plot, arising from uncertainty in solution conductivity, would span only 0.2% of a major axis division, and are therefore omitted. Prediction errors arising from ignoring the effect of the electrodes and magnet on ultrasonic propagation and field patterns were not considered here, which may account for some of the discrepancy between prediction and measurement at larger wavelengths.

Next, we conducted an experiment in 1 M KCl solution, but with the electrodes placed at a location where the magnetic field strength is no more than 0.04 $T$ (about 5.5 $cm$ from the edge of the magnet gap). This brought the electrodes to a distance of about 7 $cm$ from the transducer, leading to an expected TOF of roughly 47 $\mu s$. As shown at the bottom of Figure 6, we observed no signal at the expected arrival time, indicating a successful negative control. The number of cycles was shortened from 30 to 10 so that any Lorentz signal would not coincide with the ringing caused by the electrode impedance. The amplitude and dynamic response of the parasitically coupled signal are about the same as for the case shown in Figure 5, lending credence to the hypothesis that the electrode-solution interface impedance shapes this signal.

Figure 7 compares measured Lorentz current amplitude with prediction across three parameters: acoustic frequency, acoustic amplitude, and solution conductivity. To generate these predictions, we first estimated the ultrasonic particle velocity, $\mathbf{v_{US}}$, at the location of the electrodes using a frequency-domain axisymmetric FEM simulation (*COMSOL Multiphysics 5.1*). We integrated the component of $\mathbf{v_{US}}$ normal to both the electrode surface and the magnetic field across the electrode surface, and multiplied by the magnetic field strength and solution conductivity to arrive at the prediction for the Lorentz current amplitude. The measured Lorentz current amplitude was found by dividing the voltage amplitude at the output of the AFE by the known $Z_{trans}$.

The leftmost plot in Figure 7 compares measurement with prediction over ultrasonic frequency. The predicted trend over frequency is mainly due to the bandpass electrical-to-acoustic response of the transducer, with a smaller contribution by the change in the surface integral of $\mathbf{v_{US}}$ over frequency (note that Lorentz transduction itself is frequency-independent). Measured values at higher frequencies (1.1 $MHz$ and above) are within a margin of error of predicted values. Below 1.1 $MHz$, the measured value differs from prediction by about a factor of two, possibly because of interference set up in the cavity formed by the electrodes and the magnet gap, which is greater at longer wavelengths. For simplicity, we ignored the effects of the magnet and the electrodes in finding the expected acoustic intensity at the measurement location; future work includes modeling this geometry more precisely to arrive at more accurate predictions for expected current amplitudes at lower frequencies.

The plots in the center and right-hand-side of Figure 7 compare measurement with prediction over transducer voltage and solution conductivity, respectively. Measurements lie within a margin of error of the predicted values. The close agreement between measurement and prediction as depicted in Figure 7 suggests a quantitatively correct understanding of the mechanism of generation of ultrasonically activated Lorentz current.

In conclusion, we have demonstrated the generation and measurement of ultrasonically activated Lorentz current with extreme sub-wavelength ($<\lambda/3000$) resolution at 13 cm from the ultrasonic source, with potential for even tighter localization by using focused ultrasound. We have shown Lorentz current generation at greater depth and with finer localization than prior work in this area while providing methods for current amplitude prediction that exhibit closer agreement with measurement. The proposed method of current generation can improve the operation depth and localization of biomedical applications based on RF current, and may be a pathway toward completely non-invasive electroceuticals.

We would like to thank Devleena Samanta of Professor Richard Zare's group for conductivity measurements. This work was supported by the Defense Advanced Research Projects Agency Young Faculty Award (Dr. Douglas Weber, program manager), the National Science Foundation Graduate Research Fellowship (DGE-114747), and the Department of Defense, Air Force Of-fice of Scientific Research, National Defense Science and Engineering Graduate Fellowship (FA9550-11-C-0028).

## REFERENCES

[1] S. Semenov, Phil. Trans. R. Soc. A **367**, 3021 (2009).
[2] J. Charthad, S. Baltsavias, D. Samanta, T. C. Chang, M. J. Weber, N. Hosseini-Nassab, R. N. Zare, and A. Arbabian, in *38th Annual International Conference of the Engineering in Medicine and Biology Society (EMBC)* (Orlando, 2016) pp. 541–544.
[3] B. H. Brown, Journal of Medical Engineering & Technology **27**, 97 (2003).
[4] H. Wen, J. Shah, and R. S. Balaban, IEEE Transactions on Biomedical Engineering **45**, 119 (1998).
[5] K. L. Kilgore and N. Bhadra, Neuromodulation **17**, 242 (2014).
[6] J. Yeomans, P. Prior, and F. Bateman, Brain research **372**, 95 (1986).
[7] K. Famm, Nature **496**, 159 (2013).
[8] K. J. Tracey, Scientific American **312**, 28 (2015).
[9] E. Waltz, Nature Biotechnology **34**, 904 (2016).
[10] P. Roschmann, Medical Physics **14**, 922 (1987).
[11] A. de la Zerda, Z. Liu, S. Bodapati, R. Teed, S. Vaithilingam, B. T. Khuri-Yakub, X. Chen, H. Dai, and S. S. Gambhir, Nano Letters **10**, 2168 (2010).
[12] B. K. Andrasfalvy, B. V. Zemelman, J. Tang, and A. Vaziri, Proceedings of the National Academy of Sciences of the United States of America **107**, 11981 (2010).
[13] P. H. Moose and R. F. Klaus, The Journal of the Acoustical Society of America **74**, 1066 (1983).
[14] A. Montalibet, J. Jossinet, A. Matias, and D. Cathignol, Medical & Biological Engineering & Computing **39**, 15 (2001).
[15] A. J. Campanella, The Journal of the Acoustical Society of America **111**, 2087 (2002).